\documentclass[10pt,preprint2]{aastex}
\shorttitle{Ca, Fe, and Mg in Ellipticals}
\shortauthors{Worthey et al.}

\begin{document}

\title{Ca, Fe, and Mg Trends Among and Within Elliptical Galaxies}

\author{Guy Worthey\footnote{Visiting Astronomer, Kitt Peak National
Observatory, National Optical Astronomy Observatory, which is operated
by the Association of Universities for Research in Astronomy (AURA)
under cooperative agreement with the National Science
Foundation.}, Briana A. Ingermann, and Jedidiah
Serven\footnotemark[{1}] 
}

\affil{Department of Physics and Astronomy, Washington State
University, Pullman, WA 99164-2814, USA}

\begin{abstract}

In a sample of elliptical galaxies that span a large range of mass, a
previously unused Ca index, CaHK, shows that [Ca/Fe] and [Ca/Mg]
systematically decrease with increasing elliptical galaxy
mass. Metallicity mixtures, age effects, stellar chromospheric
emission effects, and low-mass initial mass function (IMF) boost effects are
ruled out as causes. A [Ca/Fe] range of less than 0.3 dex is
sufficient to blanket all observations. Feature gradients within galaxies
imply a global Ca deficit rather than a radius-dependent
phenomenon. Some, but not all, Type II supernova nucleosynthetic yield
calculations indicate a decreasing Ca/Fe yield ratio in more massive
supernovae, lending possible support to the hypothesis that more
massive elliptical galaxies have an IMF that favors
more massive stars. No Type II supernova nucleosynthetic yield
calculations show significant leverage in the Ca/Fe ratio as a
function of progenitor metallicity. Therefore, it seems unlikely that
the Ca behavior can be explained as a built-in metallicity effect, and
this argues against explanations that vary only the Type II to Type Ia
supernova enrichment ratio.

\end{abstract}


\section{INTRODUCTION}

The determination of age and metallicity spreads of the stars that
compose galaxies is highly useful for better
understanding the formation and evolution of the content of the
universe. Galaxies that show little evidence for present-day star
formation, such as elliptical galaxies, are better laboratories for
unraveling the stellar populations than star-forming galaxies plagued
by dust, nebulae, and bursty recent star formation.

It is now well established that other galaxies, large elliptical
galaxies in particular, display chemical abundances differing that of
our own Galaxy. Nonsolar abundance ratios in elliptical galaxies were
first mentioned as a possibility by \citet{oconnell1976} and
\citet{peterson1976}, both of whom found Na and Mg features to be
strong compared to Fe and Ca features. That [Mg/Fe] was truly greater
than zero in large elliptical galaxies, and not just a side effect of
an overall lockstep enhancement, was solidified by other studies
\citep[e.g.,][]{peletier1989, wetal92, fetal92, davies93, kuntschner00,
s10}, establishing the implication that there exists an overall
enhancement of some lighter elements relative to iron.

Magnesium and other even-$Z$ elements lighter than the iron peak are
made via capture of alpha particles, starting with $^{12}$C, and going
on to include O, Ne, Mg, Si, S, Ar, Ca, and possibly Ti, where Ti
might be too close to the Fe peak elements to be included. Workers
tend to expect these elements to track one another due to their common
nucleosynthetic origin \citep{wheel89}. So Ca abundance should track
Mg abundance according to this logic.

Emerging observations of Ca indices, however, indicate that Ca tends
not to follow Mg very well. Up to now, there have been four calcium
feature indices used to measure calcium abundance in galaxies. These
are Ca4227 and Ca4455 \citep{trager98,w98}, the Ca II ``triplet'' at
$\approx$ 8500\AA\ \citep{cenarro2003}, and Ca4227$_r$
\citep{prochaska05}.

The Ca4227, Ca4455, and Ca II triplet indices suggest a mild decrease
or a nearly flat level of [Ca/Fe] with increasing [Mg/Fe]
\citep{vazdekis97,w98,trager98,hw99,saglia02, thomas03, cenarro2003,
cenarro2004, smith09}. Interestingly and uniquely, the Ca4227$_r$
index produces results to the contrary; higher obtained [Ca/Fe]
abundances follow more closely the trend of [Mg/Fe]
\citep{prochaska05}. That index was designed to cut out some CN
absorption at the expense of requiring accurate relative
spectrophotometry.

In terms of observed indices versus galaxy velocity dispersion,
Mg-sensitive indices exhibit strong correlations with galaxy velocity
dispersion (thus, roughly, galaxy mass), while a much more modest
correlation exists for Fe-peak indices \citep{w98}. The Ca II triplet
shows a modest anticorrelation between Ca and central velocity
dispersion \citep{saglia02, cenarro2003, cenarro2004}, while the trend
with Ca4227 and Ca4455 is nearly flat \citep{trager98, cenarro2004, smith09}
Sample size and velocity dispersion range from 28 to 147 galaxies and
$\sigma=40-370$ km s$^{-1}$, respectively, with no obvious trend in
results dependent on either of these sample parameters
\citep{saglia02, thomas03, cenarro2003, cenarro2004, smith09}. The
trend of Ca4227$_r$ is increasing with increasing velocity dispersion
\citep{prochaska05}, however, the authors limited their sample of 175
early-type galaxies to those with a velocity dispersion of $\sigma
\leq 230$ km s$^{-1}$ out of the full range of $\sigma=50-300$ km
s$^{-1}$.

An explanation discussed by \citet{saglia02} and \citet{cenarro2003,
cenarro2004} involves the steepening of the initial mass function
(IMF) at the low-mass end to make more dwarfs. This would increase Na
strength \citep{st1971} and decrease CaT strength, both of which
phenomena are observed in high-mass elliptical galaxies. While it
seems appealing to solve two observational effects with one
astrophysical reason, boosting the number of low-mass stars also
greatly increases the near-IR flux, reddening colors involving near-IR
passbands, and lowers near-IR M/L ratios, and drastic lower-IMF
changes are usually looked upon with ill favor on those grounds
\citep{saglia02, thomas03}.

A more mainstream explanation rests on the astrophysics of supernovae.
Generally, Type Ia supernovae produce mainly Fe-peak elements, while
Type II supernovae produce the lighter metals, and alpha capture
elements in particular. Whatever the causes [many are listed in
\citet{w98} and \citet{trager98} and include time delays, wind
scenarios, IMF changes at the high-mass end, and binarism trends], the
Fe-peak could plausibly decouple from the light elements in terms of
abundance, but the light elements should be in lockstep with each
other. Thus, the separation of the behavior of Mg and Ca is inherently
puzzling.

\citet{w98} proposed that the simplest explanation for the Ca flatness
might be due to metallicity-dependent supernova yields, where the
production or ejection of Ca is decreased in metal-rich
supernovae. This is weakly supported by the studies of \citet{ww95},
in that [$\alpha$/Ca] increases slightly with increasing metallicity,
but more recent calculations from \citet{nomoto06} show no evidence of
this trend, with Ca tracking Mg in supernova yields of all
metallicities. The latter study includes metallicities greater than
solar. With Occam's razor thus tentatively defeated, some other effect
must be operating to create a trend along the mass sequence of
elliptical galaxies, such that low-mass ellipticals, lenticulars, and
spirals have a nearly-solar mixture, while very large ellipticals
(that sit alone at the high-mass end of all galaxy morphological
types) have a light-element enhanced mixture.

Both the observational material and the models used vary quite widely
among the workers that have addressed the Ca topic. Since the
conclusions also vary, this motivates the measurement of [Ca/Fe] in a
better way, if possible. In this paper, we approach the calcium puzzle
with the Ca II Fraunhofer H and K features that are strong features in
cool stars. Elliptical galaxies are largely composed of low-mass,
cooler stars, and thus the integrated-light spectra of these galaxies
also display these strong features, leading to a nearly ideal index
for measuring calcium abundance trends if the stellar features are
indeed sensitive to Ca abundance. Although Ca H and K are subject to
absorption by the interstellar medium, elliptical galaxies contain
very little interstellar matter. The blue wavelength of the feature
ensures that ordinary G and K type stars, mostly main sequence and
warmer giants, contribute most of the light, in contrast to the CaT
8600 feature, which is dominated by red giant branch (RGB) tip and
asymoptotic giant branch (AGB) star light whose numbers, temperatures,
and spectra are all less well understood \citep{buzzoni, w94}.

In this paper we focus on a particular CaH+K index, defined in
\citet{s05} and named CaHK, and discover for the first time an
emphatic anticorrelation between CaHK and Mg $b$ index strength. We
follow in $\S$3 with an exploration of the influence of calcium
emission lines on CaHK index strengths and rule out Ca II emission
reversals as a possible culprit for calcium underabundance
trends. Admixtures of metal poor subpopulations or young
subpopulations or dwarf-enriched populations are also ruled out,
leaving abundance ratio changes as the only viable explanation. In
$\S$4, we add galaxy-internal gradient information and discuss the inferred
abundance ratio trends between galaxies and within galaxies. Finally,
$\S$5 summarizes the findings and discusses them within the context of
basic chemical evolution.

\section{MODELS AND DATA}

A grid of synthetic stellar spectra resampled to 0.5\AA\ wavelength
intervals in the range 3000\AA\ to 10000\AA\ was incorporated into the
\citet{w94} infrastructure \citep[in which evolution choices and spectral library choices are modular; cf.][]{dotter07, lee09, swt10} in order to
produce integrated-light models. This spectral grid was repeatedly
calculated with small abundance changes, for example, a 0.3-dex
increase in calcium abundance, relative to the scaled-solar mixture,
so that spectral shape changes due to elemental abundance drifts could
be calibrated. A list of 23 elements was included, and the
\citet{gs98} abundance list was used as the solar mixture.

\begin{figure}
%\epsscale{0.8}
\plotone{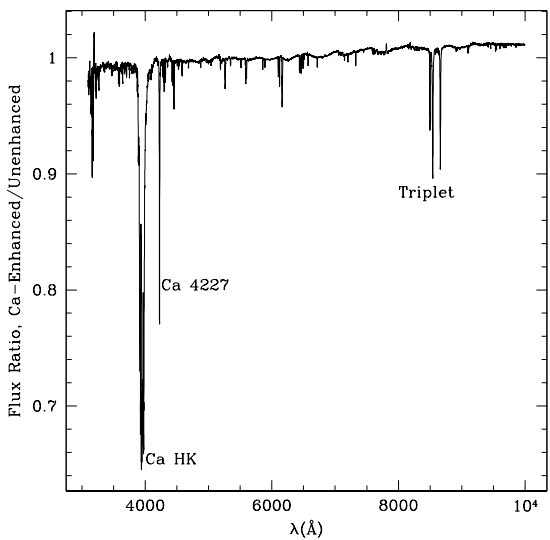}
\caption{Ratio of two model spectra of simple stellar populations of
age 15 Gyr and abundance around solar, smoothed somewhat for clarity. The model with calcium raised a
factor of two is divided by the unenhanced model. The area of the
spectrum that responds best to extra Ca is the Ca HK feature,
decreasing 30\% over a broad wavelength range. Lesser spectral responses at the
red calcium ``triplet'' and at 4227\AA\  are noted.
\label{fig:spectrum} 
}
\end{figure}

A pertinent example of this is given in Fig. \ref{fig:spectrum}, which
shows the Ca-enhanced spectrum divided by the unenhanced
spectrum. Several Ca-sensitive features become obvious. Ca does not
form molecules in stars, so the features are due to Ca I or Ca II
atomic species. It is obvious from Fig. \ref{fig:spectrum} that
the Ca H and K features give the strongest percentage response to
increased Ca abundance. Furthermore, the response is spread over a
quite broad wavelength regime compared to the other Ca-sensitive
regions of the spectrum, making measurements even easier. \citet{s05}
noted that CaHK is a factor of three more sensitive than the CaT
indices despite (in their formalism) being at much lower signal to
noise ratio (S/N). For the
same S/N, CaHK is almost an order of magnitude better.

%%% Fig 1 - ratio spectrum - ca.eps

In order to make an integrated light model population spectrum, new
stellar index fitting functions were generated \citep[first cited
in][]{poole}. The three sources of stellar spectral data were (1) a
variant of the \citet{wetal94} Lick spectra with a wavelength scale
refined via cross-correlation, (2) the MILES spectral library of
\citet{miles}, and (3) the Indo-US library of \citet{valdes}. The
Indo-US library, smoothed to mimic a Gaussian velocity dispersion with
$\sigma = 200$ km s$^{-1}$, was adopted as the fiducial set, with the
other two adjusted to fit via linear transformations. Both sets, in
general, needed more than merely zero-point shifts. In the Lick case,
a slope term was sometimes used to account for the difference in
spectra resolution. In the MILES case, a slope term was often applied
to account for what is probably a scattered light effect, which is
discussed in greater depth in \citet{miles}.

The data were then fit with multivariate polynomials in five generously
overlapping temperature swaths, then summarized in a lookup table that
gives each index strength as a function of stellar atmosphere
parameters. As described so far, the fits mimic only the run of
Galactic star index strengths with a single abundance parameter,
[Fe/H]. In order to add the sensitivity to individual elements, the
synthetic spectra were again employed to generate index responses
$\partial I/\partial $[X/R], where $X$ is the element of interest,
mostly calcium in this paper, and $R$ stands for ``generic heavy
element,'' and is employed so that the common case of Fe abundance
changes can be written as [Fe/R] and will still make logical sense.

This library of element-sensitive indices was then attached to
stellar evolutionary isochrones by choosing a \citet{sal} IMF.
The isochrone sets used in the modules are modular, but
initial testing was done with the \citet{w94} set. To produce an
integrated-light index strength, values of the flux and continuum for
each ``star'' in the isochrone, weighted by number, are added up the
isochrone, and then reformed into an index value after
summation. Finally, since we compare in the end to spectra that are
smoothed to mimic a 300 km s$^{-1}$, index corrections were
calculated using synthetic model spectra smoothed to the relevant
velocity dispersions and applied to the model indices.

For observational comparison, we use four data sources for the index
measurements. They are the \citet{graves} SDSS averages of non-LINER
elliptical galaxies and three sets of high-S/N near-nuclear spectra
\citep{san06, s10, swt10, trager2005, trager2008}. The Graves average
spectra are parceled out into bins of velocity dispersion, covering a
wide range. The CaHK index is defined in \citet{s05} and the regular
Lick indices are defined in \citet{trager98}.

The \citet{trager2008} spectra were kindly provided by the lead author
of that paper, already smoothed to 300 km s$^{-1}$. They are 12
early-type galaxies in the Coma cluster with velocity dispersions
between 41 and 270 km s$^{-1}$, including the cD galaxy NGC 4874,
taken with the Low-Resolution Imaging Spectrometer on the Keck II
10 m Telescope. This instrument uses slitlets as spectral
apertures, and the reduced spectra were weighted along the slit to
mimic a circular aperture of diameter 2.7\arcsec . The S/N in a 75 km
s$^{-1}$ bin was between 35 and 292, and one flux standard was
observed for purposes of applying a spectral shape correction. We were
also kindly provided a subset of 35 of the 98 \citet{san06} long slit
spectra, trimmed to a 4\arcsec  equivalent (at redshift $z=0.016$ for
a physical size of 0.62 kpc assuming $H_0 = 70$ km s$^{-1}$
Mpc$^{-1}$) central aperture. They were obtained at the 4.2m William
Herschel Telescope at Roque de los Muchachos Observatory using the
ISIS spectrograph and the Cassegrain Twin Spectrograph on the 3.5m
German-Spanish Astrophysical Observatory. The 35 elliptical galaxies
cover a range of velocity dispersion between 130 and 330 km s$^{-1}$
and also local density. The spectra are at a range of native
instrumental resolution between 3.5\AA\ and 6.6\AA\ (FWHM) and a
typical S/N of 50-110 \AA $^{-1}$. A few flux standards were observed
during the runs for purposes of spectra shape correction. We note that
CaHK is near the blue edge of the spectral coverage. The Serven
\citep{s10,swt10} spectra of mostly Virgo elliptical galaxies span a
range of velocity dispersion from 80 to 360 km s$^{-1}$ and were
obtained with the Cassegrain Spectrograph at the 4 m Mayall telescope
at Kitt Peak National Observatory. Nuclear portions of the long slit
spectra were extracted at an aperture of 13.8\arcsec , corresponding
to a physical distance of 1.1 kpc assuming a Virgo cluster distance of
16.47 Mpc \citep{blake09}. The S/N is over 100 per 2\AA\ bin except
for the dwarf galaxies in the sample. Dozens of flux standard star
observations of several flux standards at varying airmasses and with
both narrow and wide slits were observed nightly during the KPNO runs
for spectra shape correction. Of the various data sets, the
\citet{s10} spectra are most likely to have the most reliable
spectrophotometric shape, due to the many flux standards observed and
also because the CaHK index falls in the middle of the spectral range
of the instrument where scattered light was demonstrably negligible,
even for red objects, and camera focus issues were never observed.

The apertures sample beyond the half-light radius of each SDSS galaxy
but are more nearly nuclear for the high-S/N sets of local galaxies.
For galaxies (or average galaxies) with velocity dispersion greater
than 300 km s$^{-1}$, we computed corrections via model spectra the
same as for the model indices described above but in the opposite
direction. For galaxies with velocity dispersion less than 300 km
s$^{-1}$, extra, wavelength-dependent Gaussian smoothing was applied
to bring them to 300 km s$^{-1}$ before indices were measured.

The CaHK index definition is quite broad to capture most of the
absorption signal that is present. This broad wavelength span leads to
probable zero-point offsets between data sets due to imperfections in
recovering relative spectrophotometric shape, as discussed in
\citet{wo97}. There are only two galaxies in common between
\citet{s10} and the subset of the \citet{san06} with which we worked,
and within the possible offsets between the sets due to aperture
effects and small number statistics no mutual zero-point offset in the
CaHK index can be unambiguously shown. No other data sets have
overlap.

\begin{figure}
%\epsscale{0.45}
\plotone{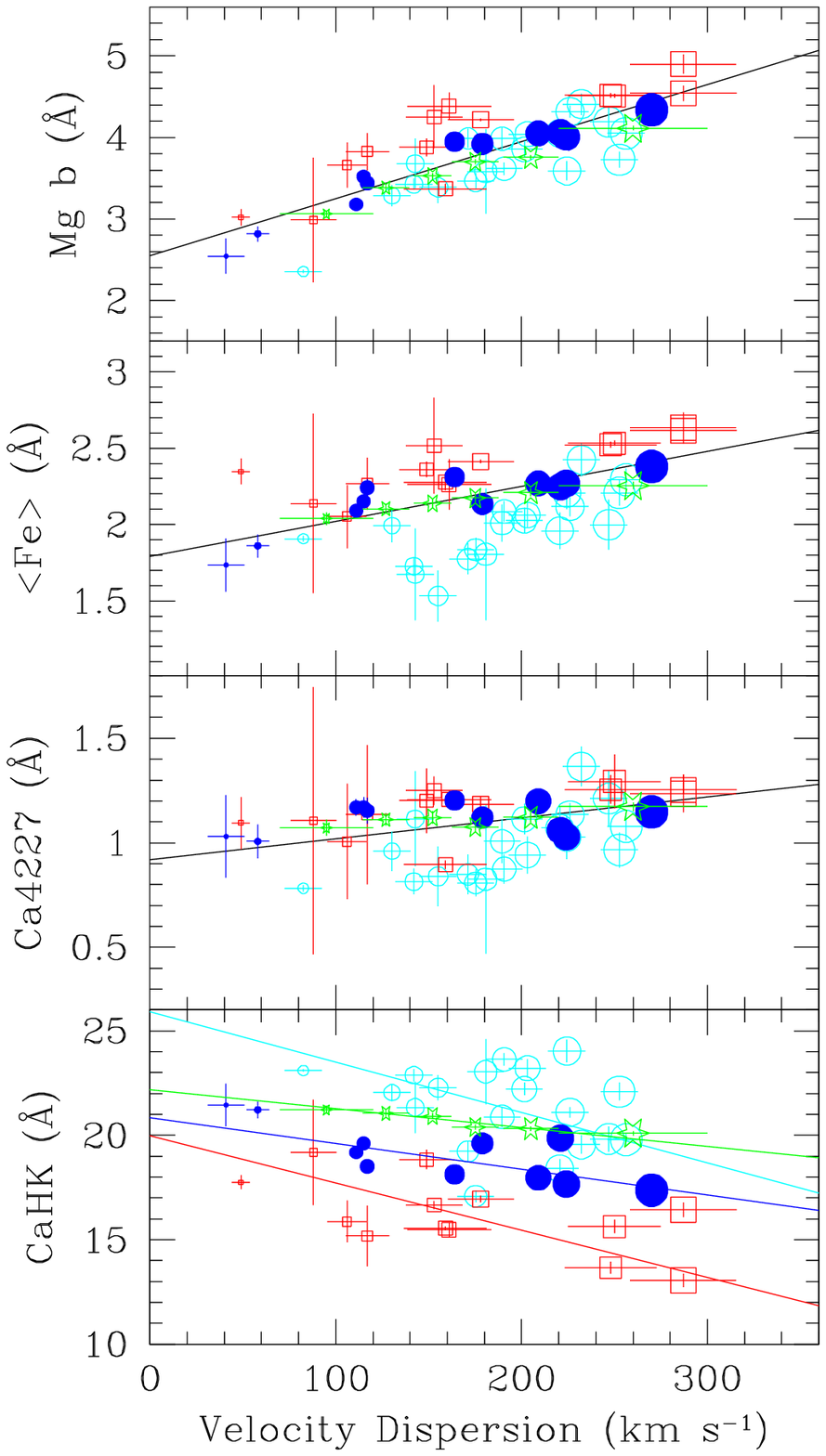}
\caption{Four feature strength indices vs. galaxy central velocity
dispersion. Data from \citet{s10,swt10} (red squares), \citet{san06}
(light blue open circles), \citet{graves} (green stars), and
\citet{trager2008} (filled blue circles) are plotted with symbol size
proportional to velocity dispersion.  The lines are two error linear
fits to the combined data, except in the bottom panel, where four
lines are shown, split by data set to show systematic offsets.
\label{fig:sigma} }
\end{figure}

Some basic index behaviors against velocity dispersion are shown in
Figure \ref{fig:sigma}. The Mg $b$ index correlates strongly with
velocity dispersion, with wide range (2\AA ). Indices $<$Fe$> = 0.5
\times ($Fe5270 + Fe5335) and Ca4227 correlate at lesser
significance. The Ca HK index, in contrast, anticorrelates. The
offsets between data sets, highlighted in the bottom panel of Figure
\ref{fig:sigma} by splitting the data into the four separate sets and
fitting lines separately, are predominantly due to (relative)
spectrophotometric shape effects.  The data are fit with the {\em
  FITEXY} program from \citet{nr} and the resulting two-error
least-squares fits are summarized in Table \ref{table2}.

The behaviors are as observed in the past for the three top indices,
all positive correlations, with Mg $b$ being the strongest correlation and
Ca4227 the weakest. The CaHK slope, however, is negative in all four data
sets. The zero point shifts we take in stride for this work since our
conclusions are not affected by them, or even by modest slope
differences, and we have not tried to aperture-correct the sets. (Such
corrections would be stronger for Mg $b$ and $<$Fe$>$ and weaker for
CaHK, based on observed gradient strengths.)

%%% Fig 2 - vd.eps

\begin{deluxetable}{lrr}
\tablewidth{0pt}
\tablecaption{Fits Versus Velocity Dispersion \label{table2}}
\tablehead{\colhead{Diagram} & 
\colhead{Intercept (\AA )} &
\colhead{Slope (\AA\ per km s$^{-1}$)}  }

\startdata
Mg $b$, combined   & $ 2.55 \pm 0.07$ & $ 0.0070 \pm 0.0003$ \\
$<$Fe$>$, combined & $ 1.79 \pm 0.04$ & $ 0.0023 \pm 0.0002$ \\
Ca4227, combined   & $ 0.92 \pm 0.04$ & $ 0.0010 \pm 0.0002$ \\
CaHK, KPNO         & $19.99 \pm 0.32$ & $-0.0226 \pm 0.0018$ \\
CaHK, SDSS         & $22.19 \pm 0.43$ & $-0.0091 \pm 0.0025$ \\
CaHK, Keck         & $20.83 \pm 0.24$ & $-0.0123 \pm 0.0012$ \\
CaHK, Herschel     & $25.90 \pm 0.12$ & $-0.0241 \pm 0.0006$ \\
\enddata
\end{deluxetable}

%Mg $b$, Combined   & $ 2.5483 \pm 0.0662$ & $ 0.0070 \pm 0.0003$ \\
%$<$Fe$>$, Combined & $ 1.7877 \pm 0.0405$ & $ 0.0023 \pm 0.0002$ \\
%Ca4227, Combined   & $ 0.9225 \pm 0.0435$ & $ 0.0010 \pm 0.0002$ \\
%CaHK, KPNO         & $19.9870 \pm 0.3177$ & $-0.0226 \pm 0.0018$ \\
%CaHK, SDSS         & $22.1860 \pm 0.4271$ & $-0.0091 \pm 0.0025$ \\
%CaHK, Keck         & $20.8276 \pm 0.2350$ & $-0.0123 \pm 0.0012$ \\
%CaHK, Herschel     & $25.8962 \pm 0.1209$ & $-0.0241 \pm 0.0006$ \\

\begin{figure}
\epsscale{1.0}
\plottwo{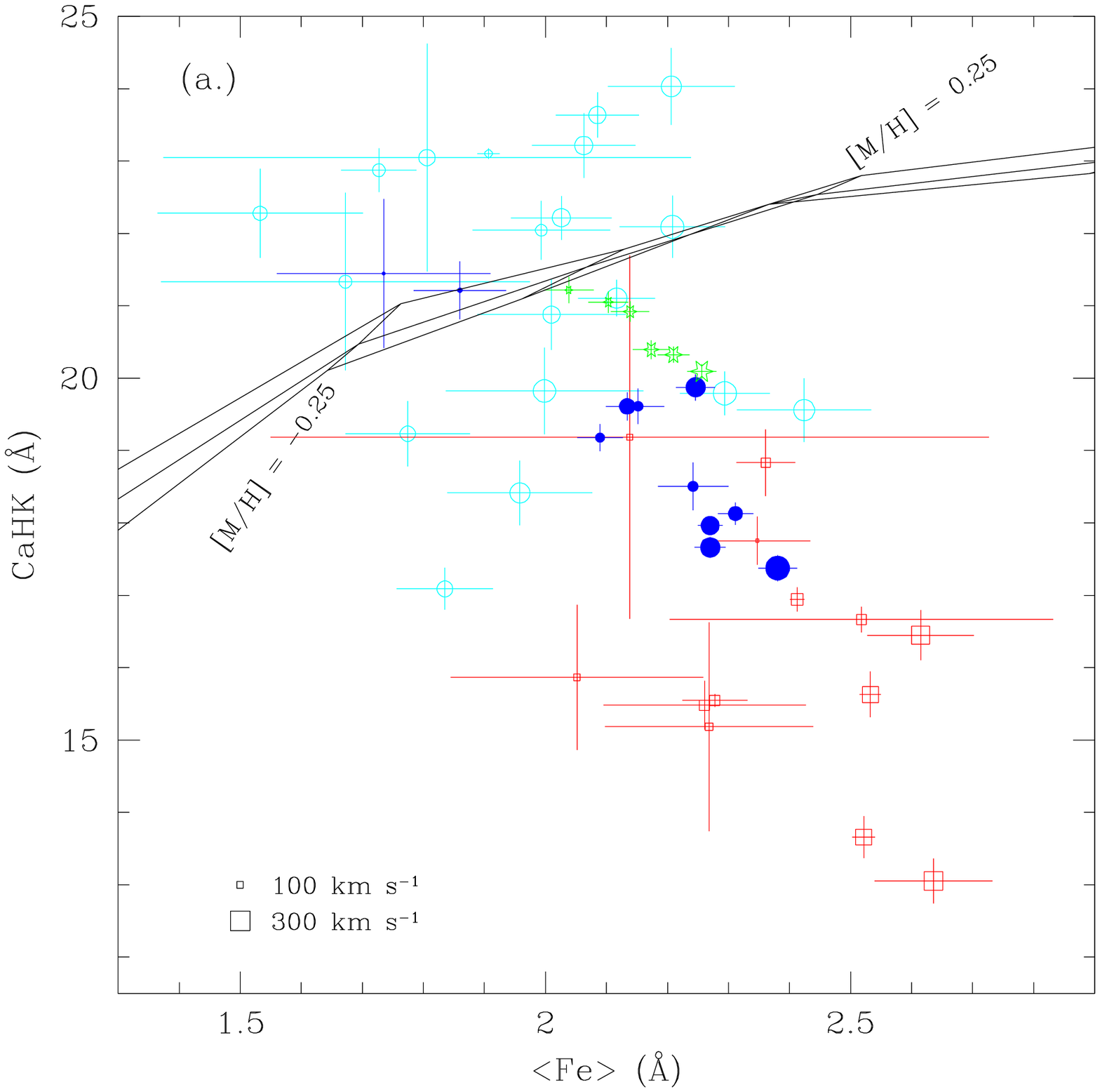}{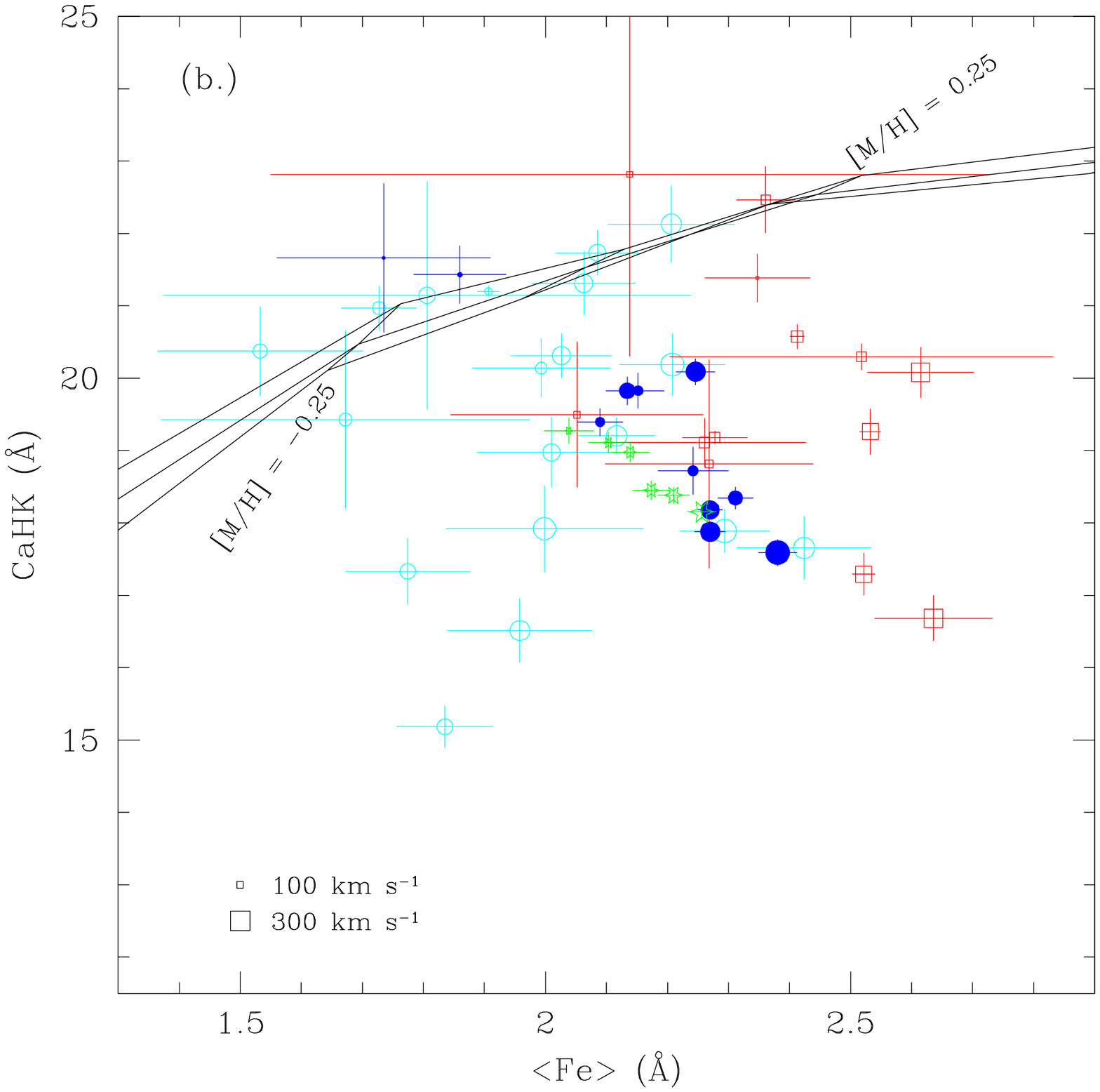}

\caption{CaHK is shown as a function of $<$Fe$>$.
Models based on the \citet{w94} isochrones are shown, at ages
8, 12, and 17 Gyr, with isometallic lines drawn at scaled solar [M/H]
every 0.25 dex.  Observational points are drawn with symbol size
proportional to galaxy velocity dispersion.  Data from
\citet{s10,swt10} (red squares), \citet{san06} (light blue open
circles), \citet{graves} (green stars), and \citet{trager2008} (filled
blue circles) are plotted. In panel (a) the data are as
measured. In panel (b) the CaHK values have been shifted to the
average value at 250 km s$^{-1}$ from the index vs. velocity dispersion
relations of Table 1.
\label{fig:CaFe}  
}

\end{figure}

\begin{figure}
%\epsscale{0.8}
\plottwo{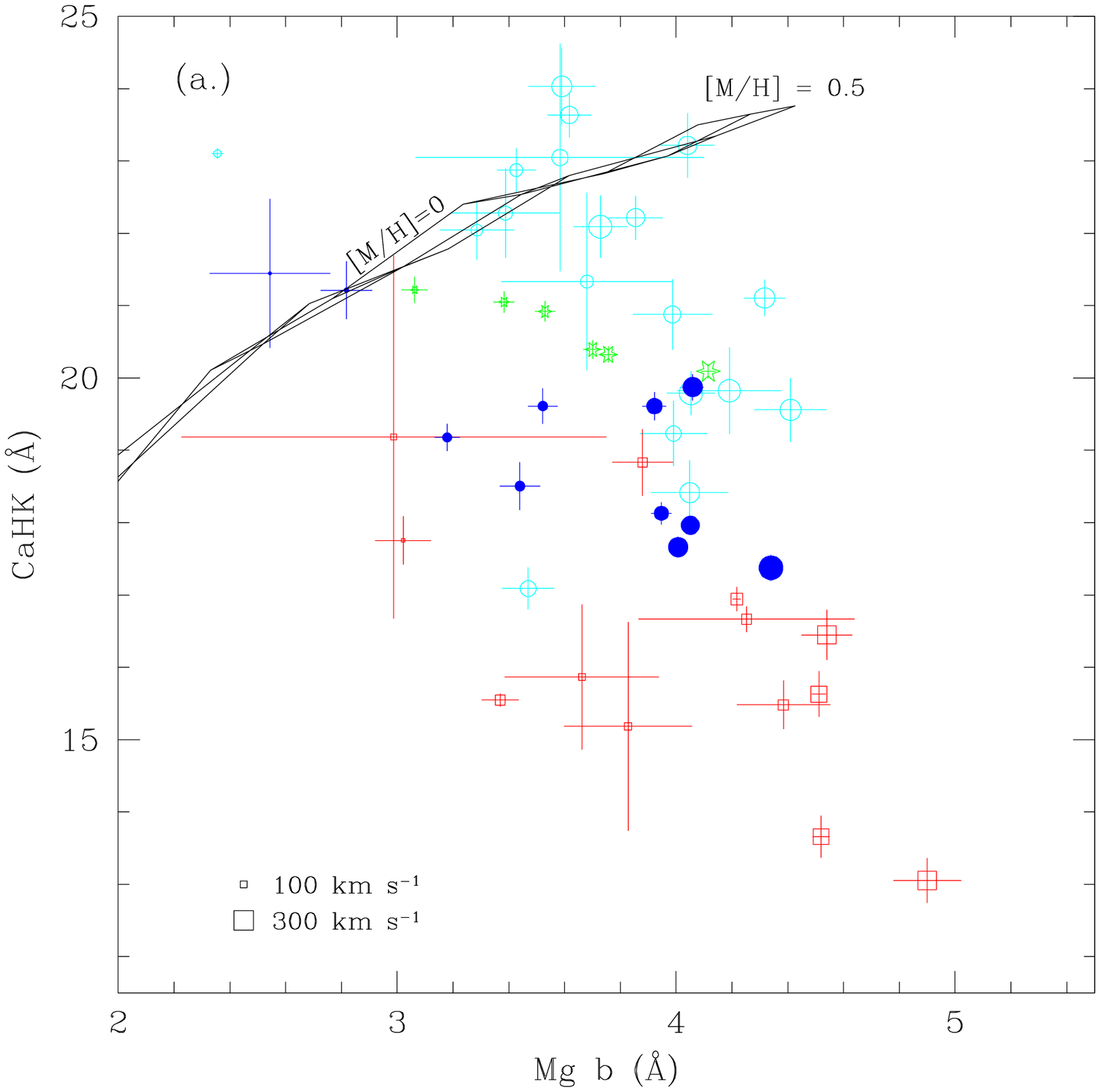}{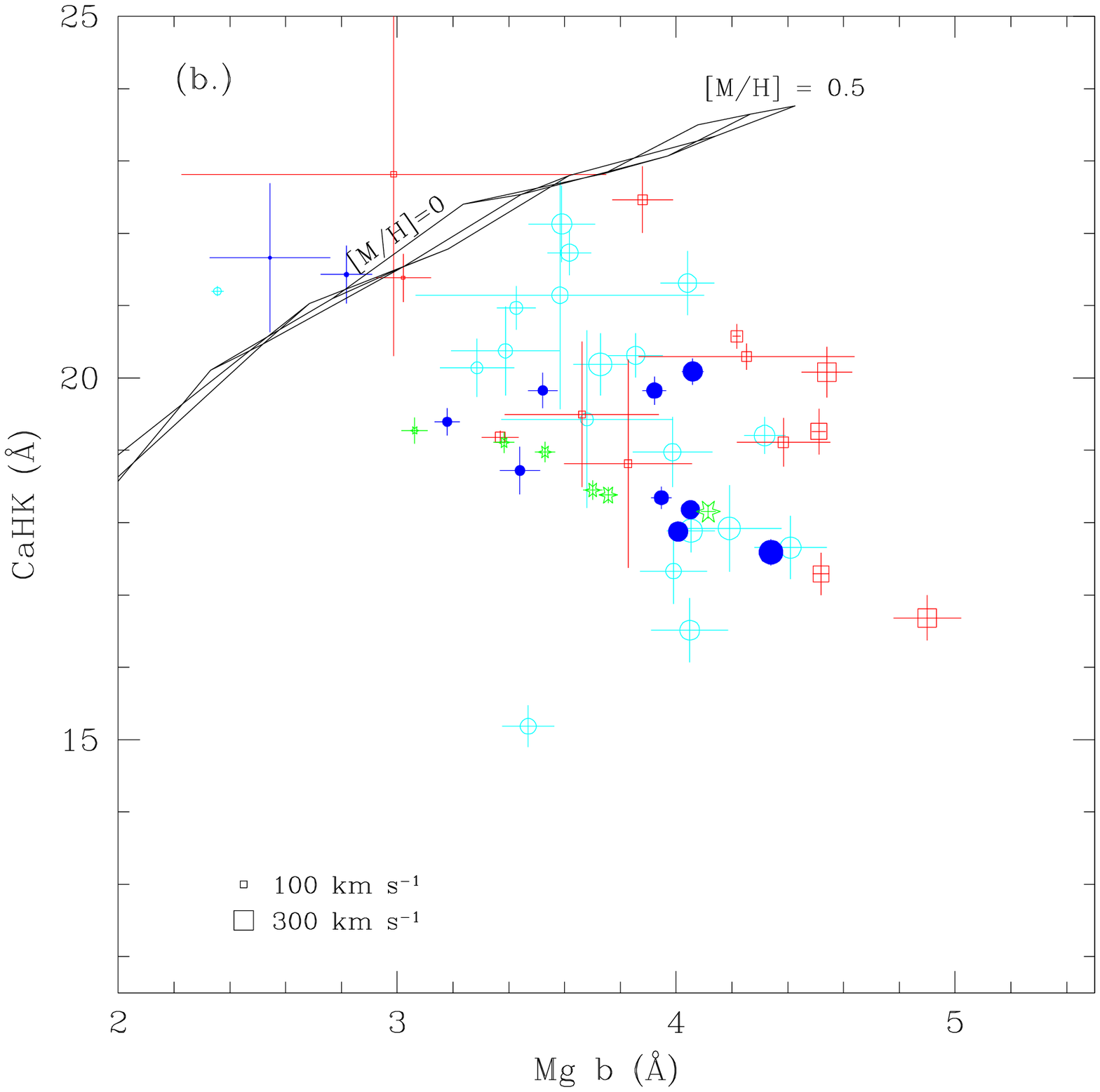}
\caption{CaHK is shown as a function of Mg $b$.
Symbols and lines are as in Fig. \ref{fig:CaFe}.
\label{fig:CaMg} 
}
\end{figure}

With these models and data in hand, we display them together in
Figs. \ref{fig:CaFe} and \ref{fig:CaMg}, arraying the CaHK index
against the $<$Fe$>$ index in Fig. \ref{fig:CaFe} and against the Mg
$b$ index in Fig. \ref{fig:CaMg}. A second panel is provided in each
case, with deltas applied to the data to make the four data sets
agree with the average value at a velocity dispersion of 250 km
s$^{-1}$. As in quite a few such index-index plots \citep{w98}, the
galaxies follow a trajectory significantly skewed from Galactic
expectations, indicating systematic drifts in abundance ratios away
from Galactic trends. In this case, the sense is that the stronger
lined elliptical galaxies have a weaker calcium index. Holding age
fixed, stronger metallic lines imply higher metallicity. Furthermore,
in the case of Mg $b$, Mg $b$ correlates with velocity dispersion
strongly \citep{bbf}, implying that it is the larger galaxies that
have the weakest CaHK index. This can be seen via the symbol size
coding in Figs. \ref{fig:CaFe} and \ref{fig:CaMg} well; except,
perhaps, for the Herschel data set, the small galaxies lie near the
models while the large ones are far away.

Iron indices have a weaker correlation
with velocity dispersion, indicating that larger ellipticals have only
slightly more Fe than small ones.  That trend is weak, and, indeed, it
is still in the realm of possibility that age systematics might be
able to drive the Fe trend.  Youth drives the stellar population left
and down in both of these plots, as seen in the model grid lines.

%%% Fig 3 - A cahkfe.eps and B cahkfe2.eps

%%% Fig 4 - A cahkmg.eps and B cahkmg2.eps

\section{EXPLANATIONS FOR CALCIUM INDEX BEHAVIOR}

From the models, it is clear that gross age and metallicity effects
cannot account for the diving CaHK index since mixtures of various
simple stellar population add approximately like vectors {\em within}
the model grid, but the data points lie outside the grid, even if one
were to shift the grid artificially. This leaves four possibilities. One,
chromospheric emission \citep{wb57} may operate with sufficient
strength to be an issue. Two, the wavelength difference between CaHK
around 3900\AA\ and the 5100\AA\ to 5400\AA\ region where Mg $b$ and
the Fe indices lie may leave the door open for a small fraction of
metal-poor subpopulation to dilute the CaHK index, with the more
metal-rich, larger galaxies having less blue flux and hence more
fill-in with the weak-lined, blue metal-poor subpopulation. Three, an
enriched dwarf population within the galaxy creates a shallower IMF on
the low-mass end and weakens the CaHK index as it might for the Ca II
triplet feature. Four, the Ca abundance relative to Fe and Mg is
changing, diluting with increased galaxy size, with implications for
supernova enrichment of the galaxies.

\subsection{Chromospheric Emission}

Within the centers of the H and K lines, emission spikes are often
present for stars of type G0 and later. This was first discovered by
\citet{se1913} and was followed up by extensive literature.
\citet{wb57} and \citet{wilson59} found the logarithms of the widths
of these emission lines to be linearly correlated with luminosity,
independent of spectral type, although stars of a given luminosity
could have a wide range of intensities of the emission. This emission
describes general chromospheric activity in stars \citep{wb57} and is
of interest because it results in weaker Ca H and K equivalent width
index measurements. We cannot use the previous work directly because
we need to judge its effects on the integrated light of stellar
populations in the particular index that we are using.

Therefore, we proceeded to estimate the {\em maximum} amount of
emission and its effect on the CaHK index by examining an
observational spectra library, choosing that of \citet{valdes} due to
spectral resolution and wavelength coverage. For purposes of this
exercise, the effective temperatures of the stars were adopted right
from the library. Giants of various temperatures, and excluding
metal-poor or chemically peculiar stars, were visually examined for
emission in the cores of the lines. If emission was spotted, a
hand-drawn guess for the true depth of the core was inserted into the
spectrum, and the CaHK index measured both with and without the
emission spike.

%%% Fig 5 - caspec.eps
\begin{figure}
%\epsscale{0.8}
\plotone{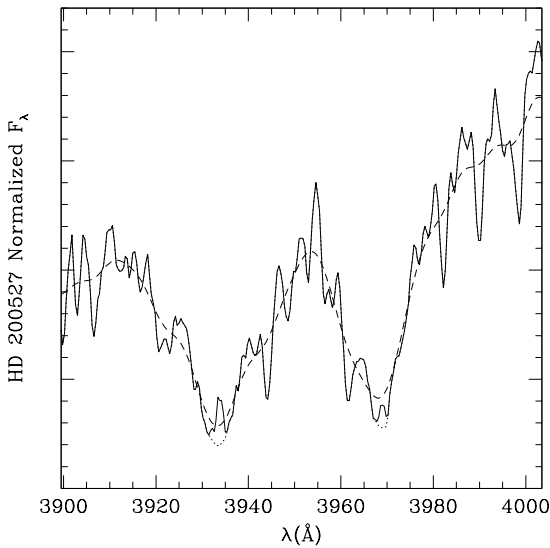}
\caption{Representative set of spectra showing Ca H
($\lambda\;3969$) and K ($\lambda\;3934$) features. The plot limits
are precisely the index passband of the CaHK index definition, while
the pseudocontinua passbands are outside the limits. The solid
line is the spectrum from the Indo-US library of \citet{valdes} for M
giant HD 200527.  The dotted portions of the spectrum
show the altered Ca H and K lines to remove emission.
The dashed line is the
latter spectrum smoothed to mimic a Gaussian velocity dispersion with
$\sigma = 200$ km s$^{-1}$.
\label{fig:caspec}
}
\end{figure}

\begin{figure}
%\epsscale{0.8}
\plotone{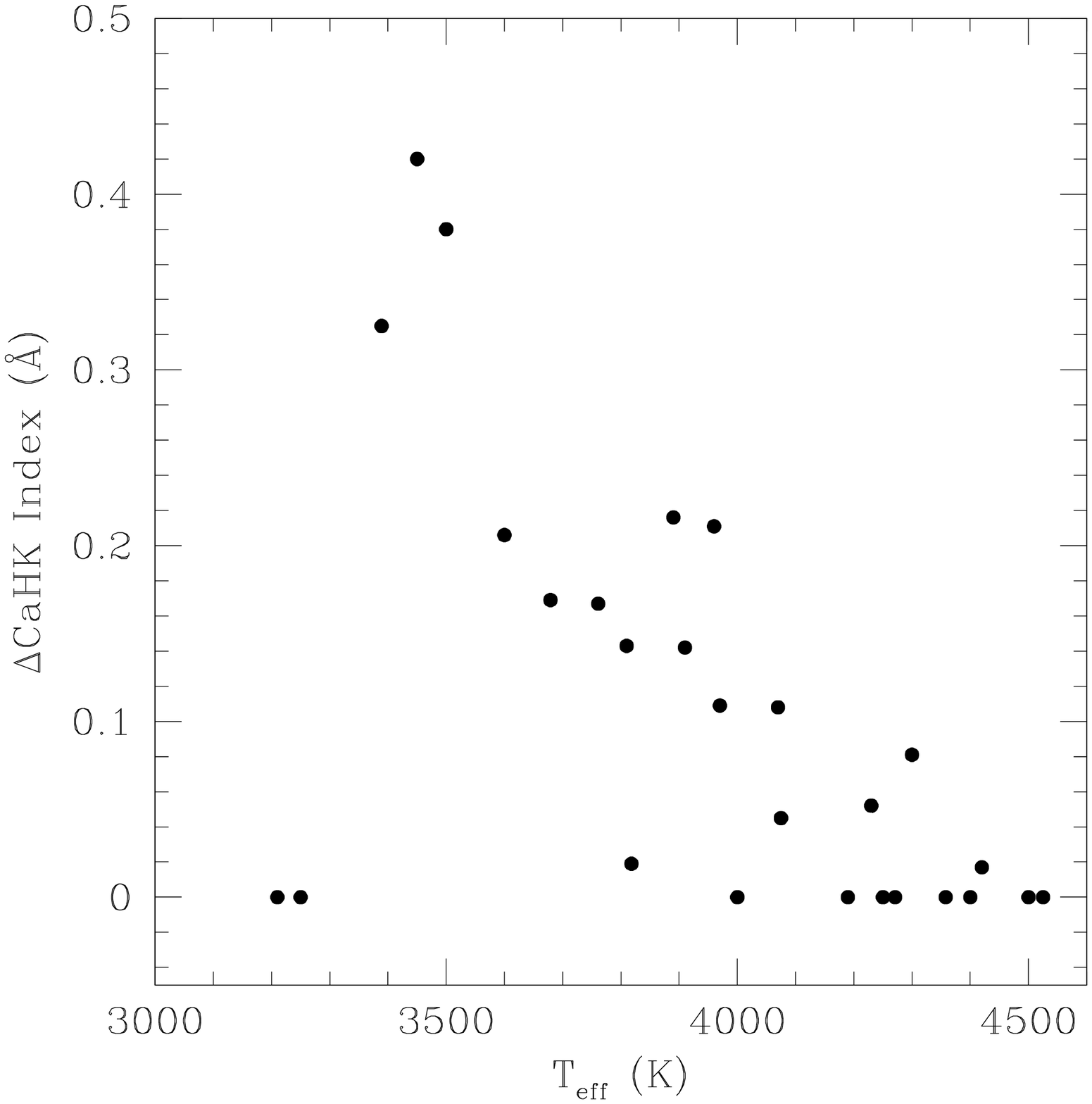}
\caption{Trend of $\Delta$CaHK = CaHK(corrected) $-$
CaHK(uncorrected) vs. $T_{\rm eff}$
for cool metal-rich giant stars selected from the Indo-US library of \citet{valdes}.
\label{fig:catest}
}
\end{figure}

Figure \ref{fig:caspec} shows the spectra, before and after, of a star
with strong emission, HD 200527. After the star is ``corrected,'' it
is smoothed to a galaxy-like velocity dispersion (200 km s$^{-1}$)
that serves as the standard for the index system. Figure
\ref{fig:catest} shows the results for a collection of K and M giants,
specifically $\Delta$CaHK = CaHK(corrected) $-$
CaHK(uncorrected). While, as expected, there is quite a lot of scatter
in the intensity of the emission, there is also a fairly well-defined
maximum edge. We examined many stars hotter than 4500 K, but none of
them had any convincing CaII emission. The late-type giants examined here
have CaHK index values of order 30\AA\ .

%%% Fig 6 - catest.eps

In order to complete the test, we approximated the $\Delta$CaHK
correction to be a line that is 0.5 \AA\ at 3000 K and 0.0 \AA\ at 4500
K and hotter, and then applied this function within the population
models of \citet{w94}. At ages 3, 6, and 12 Gyr, the $\Delta$CaHK
values were 0.015\AA\ , 0.016\AA\ , and 0.024\AA\ , respectively. The
small magnitude of these shifts results from the cool onset of the CaII
emission phenomenon, and the rapidly falling continuum levels in these
stars, so that these stars have very minimal contribution at the blue
wavelengths of Ca H and K. Definitively, then, the hypothesis that
CaII emission has anything to do with the peculiar morphology of
Figs. \ref{fig:CaFe} and \ref{fig:CaMg} is ruled out.

\subsection{Metal-Poor Subpopulations}

The second hypothesis, that a varying ratio (in light at 4000 \AA) of
metal-poor to metal-rich subpopulations can account for the galaxy
trend can also be rejected. Mixtures of populations add
approximately like vectors within the model grid. Therefore, one is
generally trapped within the confines of the model grid when mixing different
populations together, the grids in Figs. \ref{fig:CaFe} and
\ref{fig:CaMg} extend between [Fe/H] $= -2$ and [Fe/H] $= +0.5$ and
lie on a roughly linear, roughly orthogonal trajectory. There is no way
to produce the galaxy trend with inventive metallicity mixtures, with
the caveat that the fairly large wavelength separation between Mg $b$
and $<$Fe$>$ in the green and CaHK in the violet might possibly
introduce a nonlinearity.

\begin{figure}
%\epsscale{0.8}
\plottwo{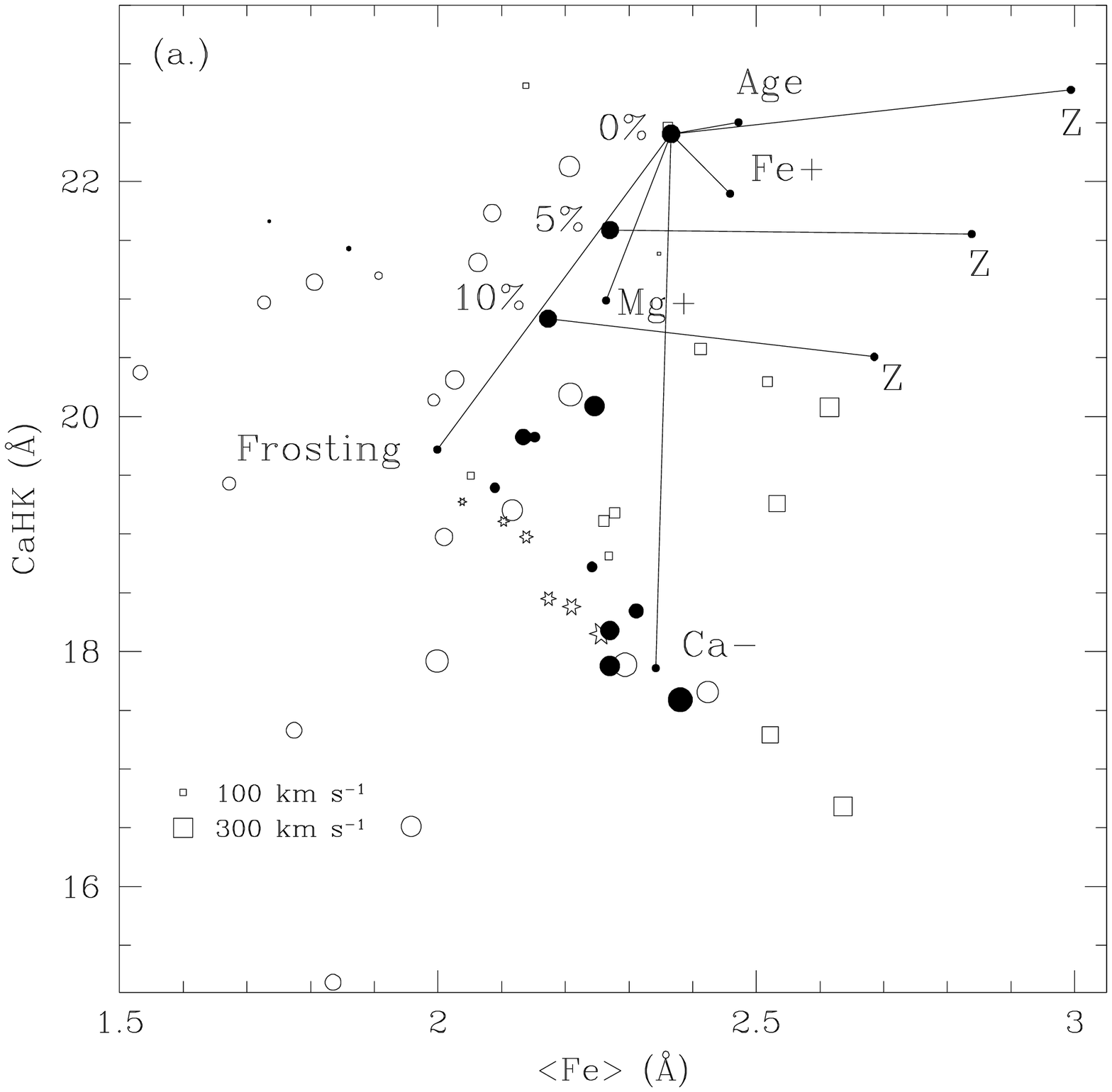}{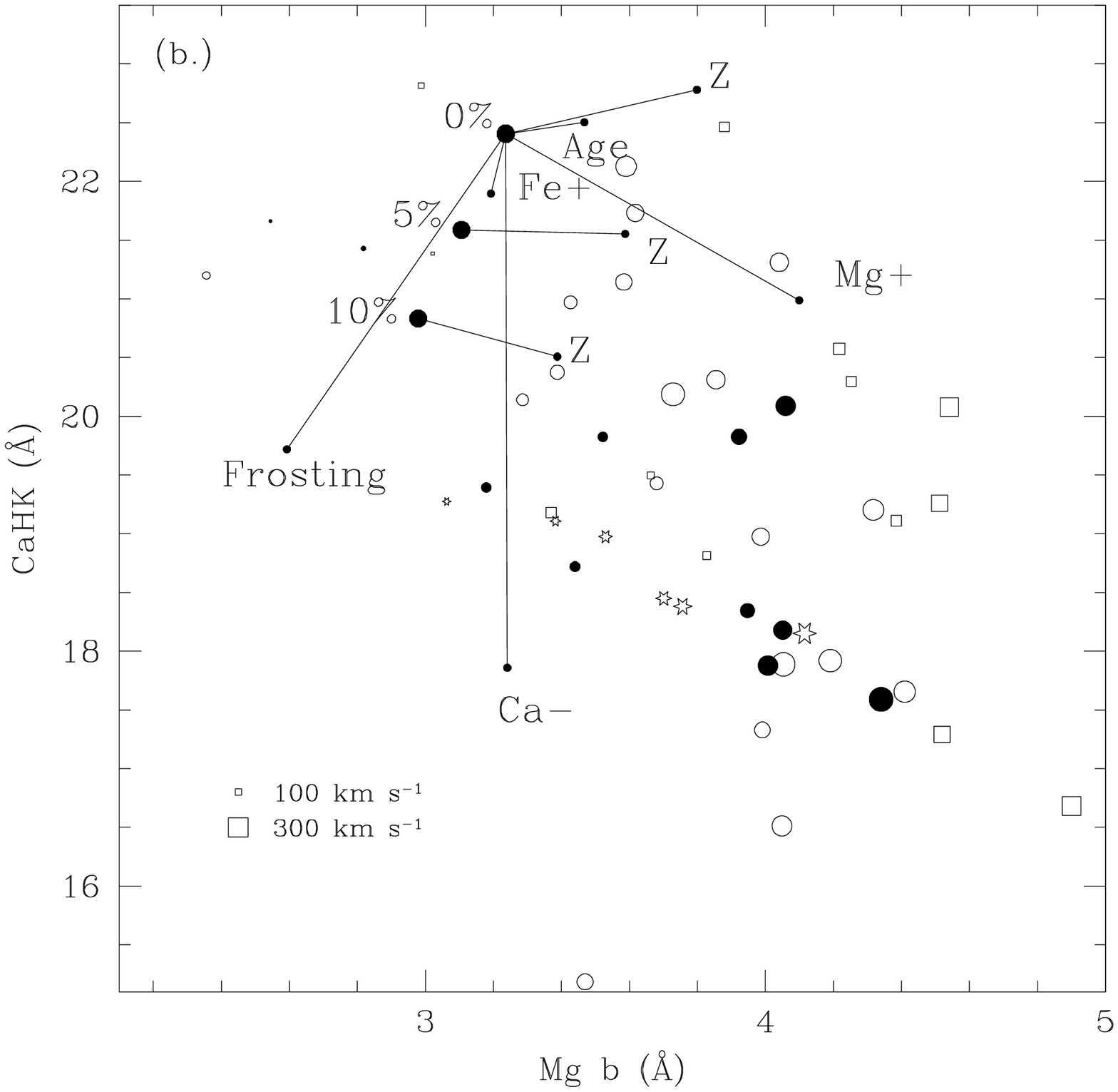}
\caption{Modeling results in the CaHK vs. $<$Fe$>$ (panel (a)) and Mg $b$
(panel (b)) diagrams are shown. 
The shifted versions of the data are repeated from previous figures,
sans error bars. Three heavy dots mark models of single-age populations,
that age being 8 Gyr. The 0\%, 5\%, and 10\% sequence refers to the
fraction by mass of an
additional metal-poor component of [M/H] = $-1.5$. The lighter dots
indicate tweaks from that initial starting model. The ones marked
``Z'' refer to increasing lockstep heavy element abundance by 0.25
dex. In the case of the metallicity-composite models, only the
metal-rich, dominant component changes its abundance. The other labels
have the following meanings. ``Age'' means an age increase from 8 Gyr
to 12 Gyr, ``Frosting'' is a 10\% by mass addition of a 1 Gyr age,
solar abundance component, ``Ca$-$'' is a drop of [Ca/R] of $-0.15$,
``Fe+'' is an increase of [Fe/R] of +0.05, and ``Mg+'' is an increase
of [Mg/R] of +0.2.
\label{fig:vectors}
}
\end{figure}

To investigate this, we concoct some more complex models in
Fig. \ref{fig:vectors}. The sequence marked 0\%, 5\%, and 10\% in
Fig. \ref{fig:vectors} is a two-metallicity composite stellar
population with the mass fraction of the metal-poor component ([M/H]
$=-1.5$) as labeled. The age is 8 Gyr for both components, and the
metal-rich component starts at [M/H] = 0 and extends to [M/H] = 0.25
at the end of the line segment marked with the letter ``Z.'' The
phenomenon of note is the slight twisting of the three line segments
as the metal-poor fraction increases. See also \citet{swt10}, who
explore indices that span even further in wavelength.

The twisting is clockwise, and that is in the correct sense to make
the model sequences flatten or even decrease with increasing
metallicity. The physical reason is that metal-poor populations,
dominated by the main sequence stars at this wavelength, are
weak-lined: as the metal-rich component drifts yet more metal-rich,
its fractional light contribution in the blue decreases dramatically,
giving more weight to the metal-poor component. There are two reasons,
however, not to take this very seriously. First, the model would imply
that more metal-rich galaxies have a more broad abundance
distribution. There is no evidence for this at less than solar [M/H]
\citep{m31}, and, one would think, little motivation to demand such a
behavior at higher abundance. Second,
metal-poor fractions as high as 10\% are probably ruled out by other
observations with less than 5\% preferred \citep{gdwarf} and
this leaves the model twists at a very mild level.

%%% Fig 7 - A cahkfe3.eps and B cahkmg3.eps

Figure \ref{fig:vectors} also shows the model vectors for bulk age
increase and lockstep abundance increase, for basic orientation. It
also shows three vectors for Ca, Fe, and Mg abundance mixture tweaks
that are useful for the discussion that follows.

\subsection{Dwarf Enrichment}

The boosted lower IMF hypothesis of \citet{saglia02} and \citet{cenarro2003} can
be addressed with the CaHK index as well. Models that include lower
main-sequence stars down to 0.1 $M_\sun$, of age 12 Gyr and of solar
composition were computed. These models used \citet{w94} isochrones
due to the availability of low-mass stars in them. The IMF was assumed
to be a power law. Slopes of 2.35 \citep{sal} and 5.35 produced CaHK
index values of 22.48\AA\ and 21.04\AA\ , respectively. However, for
this $\Delta$CaHK $\sim$ 1.5\AA , the same two IMF slopes gave $V-K$
colors of 3.45 mag and 4.99 mag, respectively. Dust free elliptical
galaxies do not have colors much redder than $V-K \approx 3.6$ mag. To
get a $\Delta$CaHK from $\sim$0.5\AA\ up an order of magnitude to the
$\sim$5\AA\ needed to match the observations would require IMF slopes
much greater than 5.35. That seems impossible, given the strong
infrared color constraint, even if the lower-mass end of the IMF was
tapered and fine-tuned to decrease the number of cool dwarfs. 

The more modest proposal of \citet{vdc}, who consider slopes of 2.35,
3.0, and 3.5 with the same 0.1 $M_\odot$ lower cutoff and compare to
Wing-Ford and Na I features in the red would probably not violate
photometric constraints and would be consistent with our
observations. We note that those authors did not consider the
hypothesis of enhanced Na abundance, correction for which
would drive their results toward more Salpeter-like slopes.

\subsection{Element Ratio Trend}

The explanation that remains is that the Ca
abundance is changing systematically from smaller elliptical galaxies
to larger ones. This is similar to the trends for Mg, Na, and N,
\citep{w98} although the Ca abundance decreases instead of
increases. These trends clearly arise through chemical enrichment
processes, and primarily through supernova yields.

The magnitude of the [Ca/R] abundance spread depends on how
conservative one would like to be when interpreting
Figs. \ref{fig:CaFe}, \ref{fig:CaMg}, and \ref{fig:vectors}. If the
model-observation separation and $>$10\AA\  CaHK index spreads are to
be taken at face value, then the [Ca/R] variation must be at least 
$-0.15$ dex for the weakest-lined galaxies, as illustrated by the
vertically-down Ca vector in Fig. \ref{fig:vectors}. Naturally, the
trends in [Fe/R] and [Mg/R] must also be taken into account, and those
elements do affect the CaHK strength, by almost 3 \AA\ for a [Mg/R]
shift of 0.2, sufficient to explain the Mg-$\sigma$ trend. At face
value, that is not sufficient to avoid invoking a [Ca/R] change. (Note
that a positive $\Delta$[Fe/R] implies a negative $\Delta$[Ca/Fe].)

If one, not unreasonably, would rather put zero faith in the
observational and model zero points, then one is left with relative
change vectors, and only the massively-average SDSS data set should be
compared with relative model change vectors. This is enabled in
Fig. \ref{fig:vectors}, where, looking only at the SDSS data and
giving complete zero-point freedom to the models, the vertical change
can be explained with the 0.2 Mg enhancement, with the need for an
[Fe/R] boost of about twice what is illustrated; about 0.1 dex in
[Fe/R] (or [Ca/Fe] $ = -0.1$) while leaving [Ca/R] unchanged. Or, one
could effect the same change with depleted Ca. Either way, and under
liberal or conservative assumptions, [Ca/Fe] does decrease from small
galaxies to large.

As regards calcium, expectations from nucleosynthesis theory are
contradictory at the moment. If we look at the work of
\citet{nomoto06} and \citet{koba06} we find that there is a strong
metallicity dependence in Na yield that has not been predicted
before. This could easily explain the high Na abundance in massive
elliptical galaxies. However, the Ca yields track Ti and Mg very well
(normalized by Fe yield), both as a function of supernova progenitor
mass and as a function of supernova progenitor initial
abundance. However, if we look at the work of \citet{ww95} and
\citet{timmes95} we find a very strong dependence of Ca yield with
supernova progenitor mass: Ca decreases relative to Fe, while Ti is
about constant and Mg increases with increasing progenitor mass, and
the variation is more than the required factor of two. This is in the
correct sense to explain our observations if IMF variation (at the
high mass end) is the root cause of the abundance trend. In this
scenario, a galaxy with higher velocity dispersion will provide star
formation environments that are biased toward higher mass star
formation. This will chemically enrich the galaxy with the higher mass
progenitor yields, and if \citet{ww95} yields are used, then it could
explain the trend we see.

The other favored explanation for the observed [Mg/Fe] trend is a
timescale argument in which a longer timescale for enrichment causes
Type Ia supernovae to contribute substantially, adding Fe-peak
elements to the mixture. However, Type Ia supernovae do not add enough
Ca or Ti or lighter elements to generate observable abundance trends
within the lighter metals. Therefore, the Ca trend is not readily
explainable in this hypothesis, because no published, detailed set of
supernova yields has sufficient metallicity dependence (initial
metallicity of the progenitor stars, that is) to torque the [Ca/Mg]
ratios into what we would infer. The Ca evidence therefore tends to
disfavor the timescale argument and favor the IMF argument.

\section{INTERIOR GRADIENTS}

\begin{figure}
%\epsscale{0.8}
\plotone{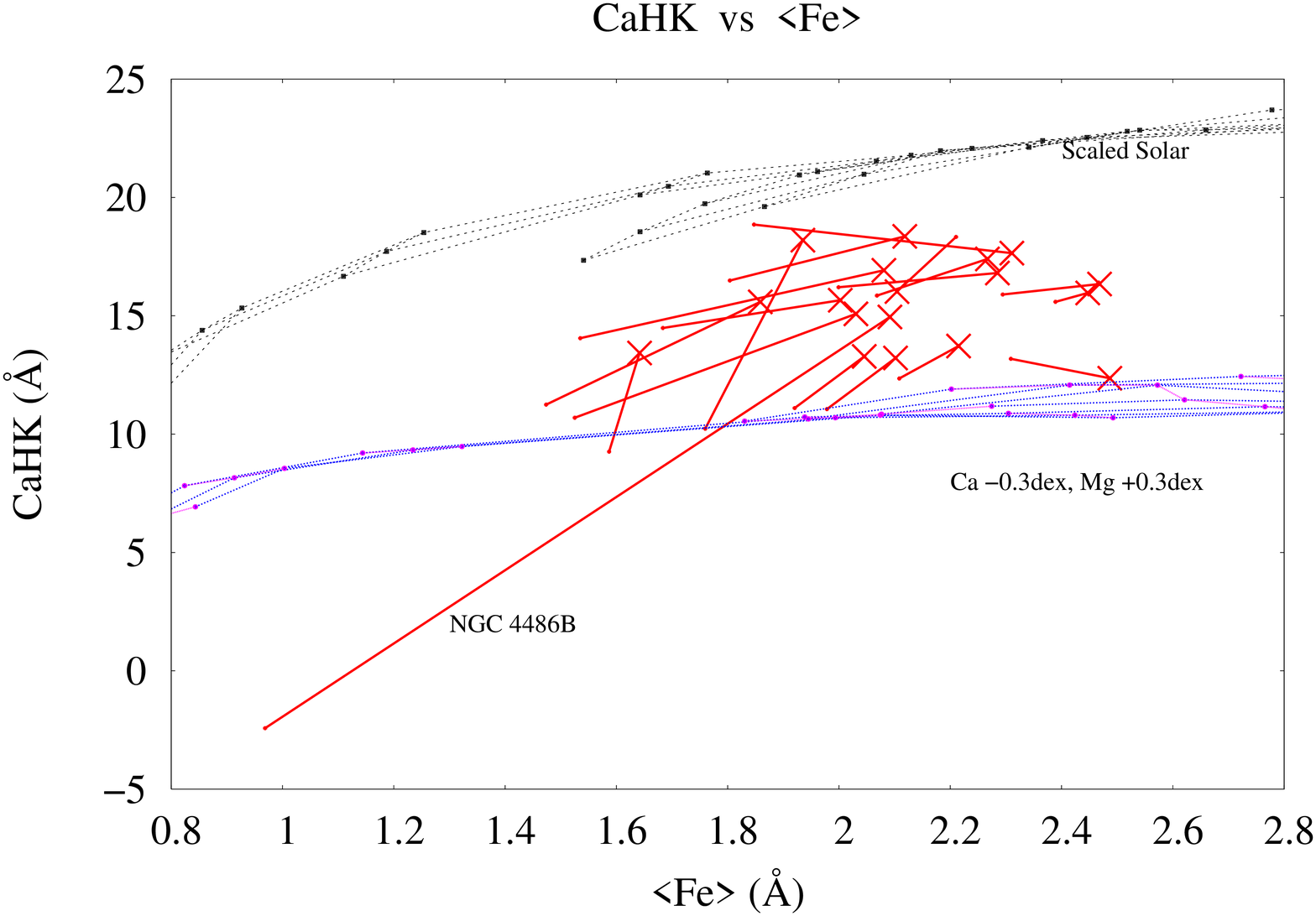}
\caption{CaHK and $<$Fe$>$ gradient trends. Using only the \citet{s10}
data set, gradient trends within each galaxy are plotted with line
segments connecting averages of the near-nuclear spectra with averages
of the outer parts of the light along the slit. The model grids contain ages 1.5, 2, 3, 5, 8, 12, 17 Gyr populations and metallicities [M/H] = $-$2, $-$1.5, $-$1.0, $-$0.5, $-$0.25, 0, 0.25, and 0.5, with no ages less than 8 Gyr plotted for metallicites less than $-$0.25. The lower set of models is identical except that, at the spectral level, 0.3 dex more Mg and 0.3 less Ca is included. This alters $Z$ in a slightly positive way, but leaves all heavy element ratios unaltered except those involving Mg or Ca. 
\label{fig:gradfe}
}
\end{figure}

\begin{figure}
\plotone{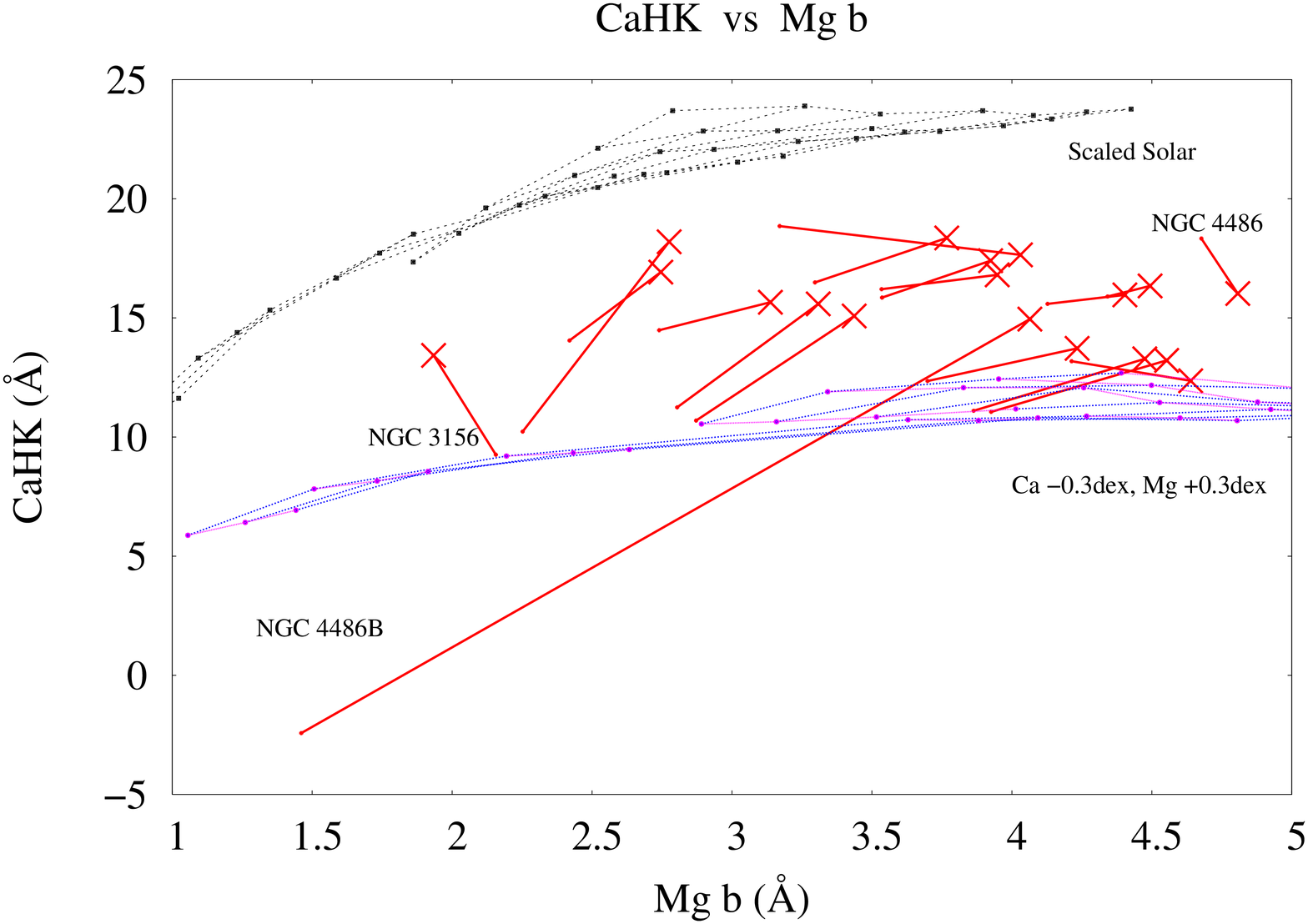}
\caption{CaHK and Mg $b$ gradient trends. Using only the \citet{s10}
data set, gradient trends within each galaxy are plotted with line
segments connecting averages of the near-nuclear spectra with averages
of the last reliable outer measurement along the slit. Models are as in Fig. \ref{fig:gradfe}. NGC 3156 is dominated by a $\sim$1 Gyr
young stellar population and also appears to buck the gradient trend
defined by the other galaxies somewhat. 
\label{fig:gradmg}
}
\end{figure}

One of our data sets contains spatially resolved indices. In
Figs. \ref{fig:gradfe} and \ref{fig:gradmg} we show an additional
model grid that has been altered such that Mg is enhanced by a 0.3
dex, and Ca is decreased by 0.3 dex, so that [Ca/Fe] $=-0.3$ and
[Ca/Mg] $=-0.6$ and [Mg/Fe] $=+0.3$. The heavy element mass fraction
is not (quite) preserved in this altered mixture, and increases a bit
because Mg is more abundant than Ca. This modest amount of alteration
is sufficient to explain even the most extreme cases. Therefore we
predict that the Ca dilution is less than a factor of two relative to
a scaled-solar abundance mixture.

%%% Fig 8 - CAHKrFe.eps

%%% Fig 9 - CAHKrMgb.eps

Figs. \ref{fig:gradfe} and \ref{fig:gradmg} also show gradient trends
from the \citet{s10} spectra, where the near-nuclear regions are
averaged and plotted as crosses, and the more distant regions are
averaged and plotted as the endpoints of trend lines. The endpoints
are the most distant points available with good data, and the end
distance is quite variable, but the main point of the figure is to
display the most reliable vector direction. With some exceptions, the
gradients tend to lie along the age-metallicity direction indicated by
the model grid, and not along a direction more indicative of strong
internal gradients in the [Ca/Fe] or [Ca/Mg] abundances. This is in
agreement with previous observations of Mg gradients \citep{wetal92}
and indicates a rather global enrichment pattern throughout the
galaxy. More recently, \citet{kuntschner06} and \citet{rawle} present
observational and model-dependent results, respectively, that describe
steady, or, on average, perhaps slightly negative Ca abundance
gradient trends with increasing galaxy radius. The data presented here
are certainly not sufficient to rule out those conclusions.

A straightforward way to homogeneously enrich the volume within a
galaxy is to have fairly effective spatial mixing during the
enrichment process. Since the process itself is almost certainly
supernova blast waves, this seems reasonable. Other mechanisms
could also contribute, such as phase mixing of the stars by dynamical
effects long after the enrichment phase is over.

% ---------------------------------------------------------------------

\section{SUMMARY AND CONCLUSION}

Four samples of elliptical galaxies that span a large range of mass
are compared to new models that feature a previously unused Ca index,
CaHK, and that are also quite flexible in displaying chemical
abundance effects on spectra and spectral indices for integrated
light. It is seen that [Ca/Fe] and [Ca/Mg] systematically decrease
with increasing elliptical galaxy Mg and Fe line strengths, and, by
implication, galaxy mass. Metallicity mixtures and age effects are
ruled out as causes due to inability to match the
observations. Stellar chromospheric emission was explored but rejected
as being about two orders of magnitude too weak to have any effect on
the observed trend. Modulations of the lower IMF increasing the dwarf
light fraction do not affect the CaHK index, and thus low-mass IMF
variations are also ruled out as a cause.

That leaves chemical abundance variation as the sole remaining
explanation. The size of the abundance shifts are modest. Less than a
factor of two in Ca dilution can easily explain the trend.

Feature gradients within galaxies imply a tendency toward a
pangalactic Ca deficit rather than a strongly radius-dependent
phenomenon. The element mixture trend may match a predicted supernova
progenitor mass dependence on the Ca yield, although this prediction
is not yet firm, apparently. If the mass-dependent Ca yield is
confirmed, then the observations strongly support a variable IMF (at
the upper end only) where more massive galaxies tend to form more of
the most massive stars. In any case, the Ca trend does not
particularly favor the scenario of timescale modulated Type II vs. Type
Ia supernova enrichment effectiveness that has received most mention
in the literature to date.

\acknowledgements 

Major funding was provided by the National Science Foundation grant
AST-0346347. The authors thank E. Toloba,
P. S{\'a}nchez-Bl{\'a}zquez, and S. C. Trager for providing their
spectral data.

% !!! note this crazy trick to decrease space between bib items!!!


\begin{thebibliography}{}\setlength{\itemsep}{-2mm}

\bibitem[Bender et al.(1993)]{bbf} Bender, R., Burstein, D., \& Faber, S. M. 1993, \apj, 411, 153
\bibitem[Blakeslee et al.(2009)]{blake09} Blakeslee, J. P., et
al. 2009, \apj, 694, 556
\bibitem[Buzzoni(1989)]{buzzoni} Buzzoni, A.\ 1989, \apjs, 71, 817 
\bibitem[Cenarro et al.(2003)]{cenarro2003} Cenarro, A.~J., Gorgas,
J., Vazdekis, A., Cardiel, N., \& Peletier, R.~F. 2003, \mnras, 339, L12
\bibitem[Cenarro et al.(2004)]{cenarro2004} Cenarro, A.~J., 
S{\'a}nchez-Bl{\'a}zquez, P., Cardiel, N., 
\& Gorgas, J. 2004, \apjl, 614, L101
\bibitem[Davies et al.(1993)]{davies93} Davies, R.~L., Sadler, E.~M.,
\& Peletier, R.~F. 1993, \mnras, 262, 650
\bibitem[Dotter et al.(2007)]{dotter07} Dotter, A., Chaboyer, 
B., Ferguson, J.~W., Lee, H.-c., Worthey, G., Jevremovi{\'c}, D., 
\& Baron, E.\ 2007, \apj, 666, 403
\bibitem[Faber et al.(1992)]{fetal92} Faber, S.~M., Worthey, G., \&
Gonz{\'a}lez, J.~J. 1992, in IAU Symp. 149, The Stellar Populations
of Galaxies, ed. B. Barbuy \& A. Renzini (Dordrecht: Kluwer), 255
\bibitem[Graves et al.(2007)]{graves} Graves, G.~J., Faber, 
S.~M., Schiavon, R.~P., \& Yan, R.\ 2007, \apj, 671, 243
\bibitem[Grevesse \& Sauval(1998)]{gs98} Grevesse, N., \& Sauval, A. J. 1998, Space Sci. Rev., 85, 161
\bibitem[Henry \& Worthey(1999)]{hw99} Henry, R.~B.~C., \& Worthey,
G. 1999, \pasp, 111, 919
\bibitem[Kobayashi et al.(2006)]{koba06} Kobayashi, C., Umeda, H.,
Nomoto, K., Tominaga, N., \& Ohkubo, T. 2006, ApJ 653, 1145 
\bibitem[Kuntschner(2000)]{kuntschner00} Kuntschner, H. 2000, \mnras,
315, 184 
\bibitem[Kuntschner et al.(2006)]{kuntschner06} Kuntschner, H., et al. 2006, \mnras, 369, 497
\bibitem[Lee et al.(2009)]{lee09} Lee, H.-c. et al. 2009, \apj, 694,
902
\bibitem[Nomoto et al.(2006)]{nomoto06} Nomoto, K., et al. 2006, Nucl. Phys. A. 777, 424 
\bibitem[O'Connell(1976)]{oconnell1976} O'Connell, R.~W. 1976, \apj,
206, 370
\bibitem[Peletier(1989)]{peletier1989} Peletier, R.~F. 1989,
PhD thesis, Univ. of Groningen
\bibitem[Peterson(1976)]{peterson1976} Peterson, R.~C. 1976, \apj,
210, L123
\bibitem[Poole et al.(2010)]{poole} Poole, V., Worthey, G., Lee,
H.-c., \& Serven, J. 2010, AJ, 139, 809
\bibitem[Press et al.(1992)]{nr} Press, W.~H., Teukolsky, S.~A.,
Vetterling, W.~T., \& Flannery, B.~P. 1992, Numerical Recipes in
FORTRAN, (2nd ed.: Cambridge: Cambridge University Press)
\bibitem[Prochaska et al.(2005)]{prochaska05} Prochaska, L.~C., Rose,
J. A., \& Schiavon, R. P. 2005, AJ, 130, 2666 
\bibitem[Rawle et al.(2010)]{rawle} Rawle, T.~D., Smith, R.~J., \&
Lucey, J.~R. 2010, \mnras, 401, 852
\bibitem[Saglia et al.(2002)]{saglia02} Saglia, R.~P., Maraston,
C., Thomas, D., Bender, R., \& Colless, M. 2002, \apj,
579, L13
\bibitem[Salpeter(1955)]{sal} Salpeter, E.~E.\ 1955, \apj, 121, 161 
\bibitem[S{\'a}nchez-Bl{\'a}zquez et al.(2006a)]{san06}
S{\'a}nchez-Bl{\'a}zquez, P., Gorgas, J., Cardiel, N., \&
Gonz{\'a}lez, 
J.~J.\ 2006a, \aap, 457, 787
\bibitem[S{\'a}nchez-Bl{\'a}zquez, P. et al.(2006b)]{miles} 
S{\'a}nchez-Bl{\'a}zquez, P., et al. 2006b, \mnras, 371, 703 
\bibitem[Schwarzschild \& Eberhard(1913)]{se1913} Schwarzschild, K.,
\& Eberhard, G. 1913, \apj, 38, 292
\bibitem[Serven et al.(2005)]{s05} Serven, J., Worthey, G., \& Briley,
M.~M. 2005, \apj, 627, 754
\bibitem[Serven(2010)]{s10} Serven, J. 2010, PhD thesis, Washington
State University
\bibitem[Serven et al.(2010)]{swt10} Serven, J., Worthey, G.,
Toloba, E. 2010, \& S{\'a}nchez-Bl{\'a}zquez, P. 2011, \aj, 141, 184 
\bibitem[Smith et al.(2009)]{smith09} Smith, R.~J., Lucey, J.~R.,
Hudson, M.~J., \& Bridges, T.~J. 2009, \mnras, 398, 119 
\bibitem[Spinrad 
\& Taylor(1971)]{st1971} Spinrad, H., \& Taylor, B.~J.\ 1971, \apjs, 22, 445
\bibitem[Thomas et al.(2003)]{thomas03} Thomas, D., Maraston, C., \&
Bender, R. 2003, \mnras, 343, 279
\bibitem[Timmes et al.(1995)]{timmes95} Timmes, F.~X., Woosley, 
S.~E., \& Weaver, T.~A.\ 1995, \apjs, 98, 617 
\bibitem[Trager et al.(2008)]{trager2008} Trager, S.~C., Faber, 
S.~M., \& Dressler, A.\ 2008, \mnras, 386, 715
\bibitem[Trager et al.(1998)]{trager98} Trager, S.~C., Worthey, 
G., Faber, S.~M., Burstein, D., \& Gonz{\'a}lez, J.~J.\ 1998, \apjs, 116, 1 
\bibitem[Trager et al.(2005)]{trager2005} Trager, S.~C., Worthey, 
G., Faber, S.~M., \& Dressler, A.\ 2005, \mnras, 362, 2

\bibitem[Valdes et al.(2004)]{valdes} Valdes, F., Gupta, R., 
Rose, J.~A., Singh, H.~P., \& Bell, D.~J.\ 2004, \apjs, 152, 251
\bibitem[van Dokkum \& Conroy(2010)]{vdc} van Dokkum, P. G., \& Conroy, C. 2010, \nat, 468, 940
\bibitem[Vazdekis et al.(1997)]{vazdekis97} Vazdekis, A., Peletier,
R.~F., Beckman, J.~E., \& Casuso, E. 1997, \apjs, 111, 203 
\bibitem[Wheeler et al.(1989)]{wheel89} Wheeler, J.~C., Sneden, C., \& Truran, J.~W., Jr.\ 1989, \araa, 27, 279
\bibitem[Wilson(1959)]{wilson59} Wilson, O.~C. 1959, \apj, 130, 499
\bibitem[Wilson \& Vainu Bappu(1957)]{wb57} Wilson, O.~C., \& Vainu
Bappu, M.~K.\ 1957, \apj, 125, 661
\bibitem[Woosley \& Weaver(1995)]{ww95} Woosley, S.~E., \& Weaver, T.~A.\ 1995, \apjs, 101, 181 
\bibitem[Worthey(1994)]{w94} Worthey, G. 1994, \apjs, 95, 107
\bibitem[Worthey(1998)]{w98} Worthey, G. 1998, \pasp, 110, 888
\bibitem[Worthey et al.(1996)]{gdwarf} Worthey, G., Dorman, B., \& Jones, L.~A.\ 1996, \aj, 112, 948 
\bibitem[Worthey et al.(2005)]{m31} Worthey, G., Espa{\~n}a, A., MacArthur, L.~A., \& Courteau, S.\ 2005, \apj, 631, 820 
\bibitem[Worthey et al.(1992)]{wetal92} Worthey, G., Faber, S.~M., \&
Gonz{\'a}lez, J.~J. 1992, \apj, 398, 69
\bibitem[Worthey et al.(1994)]{wetal94} Worthey, G., Faber, S.~M.,
Gonz{\'a}lez, J.~J., \& Burstein, D. 1994, ApJS, 94, 687
\bibitem[Worthey \& Ottaviani(1997)]{wo97} Worthey, G., \& Ottaviani,
D.~L.\ 1997, \apjs, 111, 377


\end{thebibliography}
\end{document}